# PROPOSED DESIGN FOR SIMULTANEOUS MEASUREMENT OF WALL AND NEAR-WALL TEMPERATURES IN GAS MICROFLOWS

**Varun Yeachana[1], Vikash Kumar[2], Lucien Baldas[1], Marcos Rojas-Cardenas[1], Christine Barrot[1], Ryan Enright[2], Stéphane Colin[1]\***

1. Institut Clément Ader (ICA), Université de Toulouse, CNRS, INSA, ISAE-SUPAERO, Mines-Albi, UPS, Toulouse France
2. Thermal Management Research Group, Efficient Energy Transfer (ηET) Dept., Bell Labs Ireland, stephane.colin@insa-toulouse.fr

## ABSTRACT

Gas behavior in systems at microscale has been receiving significant attention from researchers in the last two decades [1-4]. Today, there is an enhanced emphasis on developing new experimental techniques to capture the local temperature profiles in gases at rarefied conditions. The main underlying reason behind this focus is the interesting physics exhibited by gases at these rarefied conditions, especially in the transition regime. There is the onset of local thermodynamic disequilibrium, which manifests as velocity slip and temperature jump [1-4] at the wall. However, there is limited experimental evidence on understanding these aforementioned phenomena. With the advances in experimental facilities, it is today possible, at least in principle, to map the local temperature profiles in gases at rarefied conditions. Molecular tagging approach is one such technique which has shown the potential to map the temperature profile in low pressure conditions [5].

In molecular tagging approach, a very small percentage of tracer molecules are introduced into the gas of interest, referred as carrier gas. In gas flow studies, the typical tracers employed are acetone and biacetyl. These tracer molecules, assumed to be in equilibrium with the carrier gas, are excited with a source of energy at a specific wavelength, typically a laser. The excited molecules are unstable and tend to de-excite in a radiative and non-radiative manner, which is manifested as fluorescence and phosphorescence. Following the deformation with time of a tagged line permits to obtain the flow velocity. In addition, the dependence of the phosphorescence and fluorescence intensity to the gas temperature could also allow to use this technique for local temperature measurements.

The objective of this study is to develop an experimental setup capable of simultaneously mapping the wall and fluid near-wall temperatures with the final goal to measure temperature jump at the wall when rarefied conditions are reached. The originality of this setup shown in Figure 1 is to couple surface temperature measurements using an infrared camera with Molecular Tagging Thermometry (MTT) for gas temperature measurements. The bottom wall of the channel will be made of Sapphire substrate of 650 μm thickness coated with a thin film of Indium Tin Oxide (ITO). The average roughness of this ITO layer is about 3 nm. The top wall of the channel will be made of SU8 and bonded with the bottom wall with a layer of PDMS. The channel will be filled in with acetone vapor, or with a carrier gas seeded with acetone molecules. The channel height $h$ and the pressure of acetone will be chosen in order to achieve a Knudsen number of the order of 0.1 corresponding to the slip or early transition regimes. The ITO layer will be uniformly heated in order to generated within the gas a temperature gradient perpendicular to the surface, that will result in rarefied conditions in a temperature jump i.e. a difference of temperature between the wall and the gas molecules.

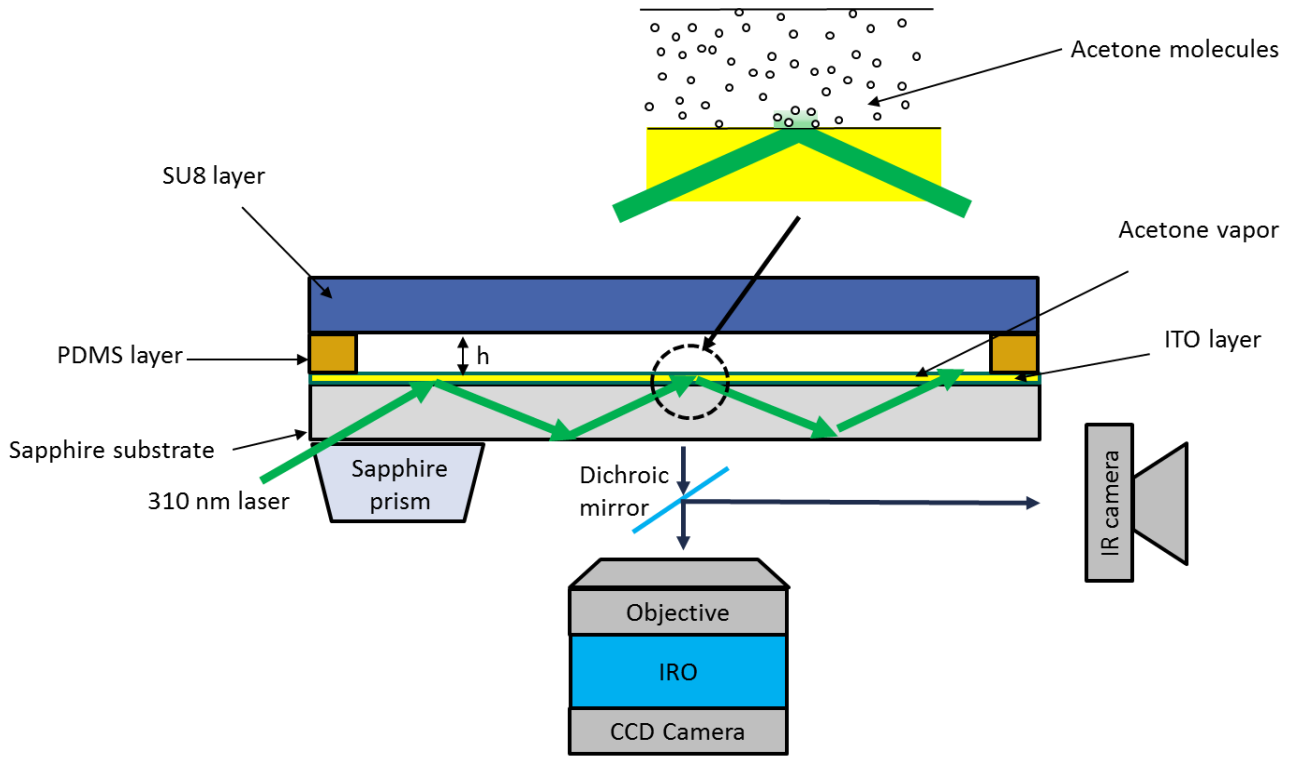

Figure 1: Schematic of the proposed experimental setup

Surface temperature measurement of the sapphire wall will be measured with the ITO layer emitting an IR signal recorded by a thermal camera. This is a non-intrusive technique to measure the wall temperature, which has been developed at Nokia Bell Labs in Dublin [6]. For measuring the temperature of the molecules close to the wall, evanescent ultraviolet (UV) waves will be used to illuminate a thin layer of fluid near the wall and to excite the acetone tracer molecules present in this layer, as shown in the zoomed view of Figure 1. A pulsed laser sheet of wavelength 310 nm will be subjected to total internal reflection (TIR) at the top end of the bottom wall (in our case a sapphire substrate with refractive index 1.75). A sapphire prism helps in bending of the light beam at the appropriate angle. To satisfy the condition of continuity of electric and magnetic fields at the boundary of reflection, waves of light are transmitted into the medium of lower refractive index (here acetone vapor with a refractive index 1.35) [9]. These waves are referred to as evanescent waves, and they exhibit an exponential decay in intensity with distance. Their presence is felt within an order 100 nm in the material of lower refractive index [7-9]. Therefore, these evanescent waves can be employed to probe the regions closer to the wall.

Evanescent wave-based studies have been previously employed by researchers, especially in multilayer Nano-PIV studies to study the near-wall behavior of liquid flows [7-9]. Also, some researchers have adopted this technique to map the temperature in liquids in regions which are few hundred nanometers from the wall [9]. In general, these techniques employ fluorescent particles. The challenge of this study will be to obtain near-wall gas temperature measurement based on phosphorescence of molecules instead of particles. The phosphorescence is captured by an Intensified relay optics (IRO) attached to the CCD camera. The IRO not only intensifies the signal of photoluminescence, but also provides with an adjustment for the gate i.e. the duration for which the signal is captured (the CCD camera is always open, but the signal is captured based on the gate value).The intensity dependence of acetone phosphorescence with temperature has been experimentally demonstrated in our previous work on MTT [5]. Capturing the phosphorescence of the acetone molecules close to the wall will give us an information of the near-wall temperature of acetone vapor.

In our previous study [5], phosphorescence emission of acetone vapor has been studied at the excitation wavelength of 310 nm, at various conditions of pressure ($p =$ 15,000 Pa, 10,000 Pa, 7500 Pa, 5000 Pa, 2500 Pa and 1000 Pa) and conditions of temperatures (20 °C, 34 °C, and 50 °C). Therefore, these conditions of temperature and pressure will be targeted in the future experiments on the proposed experimental setup. The initial aim of the study will be to carry out experiments without flow in the system. Subsequently, experiments with flow in the system will be conducted to study temperature jump. Based on the difference in temperature values obtained by analyzing both the ITO generated IR signal and the visible phosphorescent signal of acetones molecules within the Knudsen later at the wall, it should be possible to get information on the temperature jump at the wall.

**Acknowledgements**

This project has received funding from the European Union's Framework Programme for Research and Innovation Horizon 2020 (2014-2020) under the Marie Skłodowska-Curie Grant Agreement No. 643095, and from the Fédération de Recherche FERMAT (FR 3089).

**References and Citations**